%%%%%%%%%%% Plain TeX file %%%%%%%%%%%%%%%%%%%%%%%%%%%%%%%%%%%%%%%%%%%%%%

\magnification=\magstep1 
\font\bigbfont=cmbx10 scaled\magstep1
\font\bigifont=cmti10 scaled\magstep1
\font\bigrfont=cmr10 scaled\magstep1
\vsize = 23.5 truecm
\hsize = 15.5 truecm
\hoffset = .2truein
\baselineskip = 14 truept
\overfullrule = 0pt
\parskip = 3 truept
\def\frac#1#2{{#1\over#2}}

\nopagenumbers
\topinsert
\vskip 3.2 truecm
\endinsert
\centerline{\bigbfont QUANTUM ALGEBRAIC SYMMETRIES IN }
\centerline{\bigbfont ATOMIC CLUSTERS, MOLECULES AND NUCLEI}  
\vskip 20 truept
\centerline{\bigifont Dennis Bonatsos and N. Karoussos} 
\vskip 8 truept
\centerline{\bigrfont Institute of Nuclear Physics, NCSR ``Demokritos''} 
\vskip 2 truept
\centerline{\bigrfont GR-15310 Aghia Paraskevi, Attiki, Greece} 
\vskip 14 truept
\centerline{\bigifont P. P. Raychev and R. P. Roussev} 
\vskip 8 truept
\centerline{\bigrfont Institute for Nuclear Research and Nuclear Energy}
\vskip 2 truept 
\centerline{\bigrfont 72 Tzarigrad Road, BG-1784 Sofia, Bulgaria}  

\vskip 28 truept 
\centerline{\bf 1.  INTRODUCTION}
\vskip 12 truept

Quantum algebras (also called quantum groups) are deformed versions of the
usual Lie algebras, to which they reduce when the deformation parameter 
$q$ is set equal to unity. From the mathematical point of view they are 
Hopf algebras. Their use in physics became popular with the introduction 
 of the $q$-deformed harmonic oscillator as a tool for 
providing a boson realization of the quantum algebra su$_q$(2), although 
similar mathematical structures had already been known. 
Initially used for solving the quantum Yang--Baxter equation, quantum algebras
have subsequently found applications in several branches of physics, as, for 
example, in the description of spin chains, squeezed states, 
hydrogen atom and hydrogen-like spectra,
rotational and vibrational nuclear and molecular spectra, and in conformal 
field theories. By now much work has been done 
on the $q$-deformed oscillator and its relativistic extensions, 
and several kinds of generalized deformed oscillators 
and su(2) algebras 
have been introduced.  Simple and self-contained introductions to the 
quantum algebraic techniques and their applications in physics 
can be found in Refs [1--3]. 

Here we shall confine ourselves to a list of applications of quantum algebras 
in nuclear structure physics and in molecular physics and, in addition,  
a recent application of quantum algebraic techniques to 
the structure of atomic clusters will be discussed in more detail. 

\vskip 28 truept
\centerline{\bf 2. QUANTUM ALGEBRAS IN NUCLEAR STRUCTURE}
\vskip 12 truept

We give here a brief list of applications of quantum algebraic techniques 
in nuclear structure. More details and relevant references can be found 
in the review articles [1,2]. 

1) Rotational spectra of deformed and superdeformed nuclei, 
including excited (beta and gamma) bands,  have been 
described in terms of the su$_q$(2) model, in a way similar to that 
provided by the Variable Moment of Inertia (VMI) model, with the deformation 
parameter $\tau$ (with $q=e^{i\tau}$) found to correspond to 
the softness parameter of the VMI model. Through a comparison of the su$_q$(2)
model to the hybrid model, the deformation parameter $\tau$
has also been connected to the number of valence nucleon pairs 
and to the nuclear deformation $\beta$. Recently the model has been 
extended in order to be applicable to transitional and vibrational spectra. 

2) B(E2) transition probabilities have also been described in the framework
of the su$_q$(2) model. 
In this case the $q$-deformed Clebsch--Gordan coefficients are used instead 
of the normal ones. (It should be noticed that the $q$-deformed angular 
momentum theory has already been much developed.) 
The model predicts 
an increase of the B(E2) values with angular momentum, while the rigid 
rotator model predicts saturation. Some experimental results supporting 
this prediction already exist, along with theoretical predictions from 
other models towards the same direction.  

3) A generalization of the su$_q$(2) model is based on the use of the 
deformed algebra su$_{\Phi}$(2), 
which is characterized by a structure function $\Phi$. 
The usual su(2) and su$_q$(2) algebras are obtained for specific choices 
of the structure function $\Phi$. 
A two-parameter generalization of the su$_q$(2) model, labelled as 
su$_{qp}$(2), has also been successfully used for the description of 
superdeformed nuclear bands. 

4) It has been found that correlated fermion pairs coupled to zero 
angular momentum in a single-$j$ shell behave approximately as suitably
defined
$q$-deformed bosons, the pairing energies also being correctly
reproduced up to the same order, and the deformation parameter ($\tau
=\ln q$) being found to be inversely proportional to the size of the shell. 
The same system of correlated fermion pairs can be described 
{\sl exactly} by suitably defined generalized deformed bosons. Then 
both the commutation relations are satisfied exactly and the pairing energies 
are reproduced exactly. The spectrum of the appropriate generalized 
deformed oscillator corresponds, up to first order perturbation theory,
to a harmonic oscillator with an $x^4$ perturbation. 

5) A $q$-deformed version of a two dimensional toy Interacting Boson Model 
(IBM)  with su$_q$(3) overall symmetry, as well as $q$-deformed versions of 
the o(6) and u(5) limits of the full IBM have been developed. 
The $q$-deformation of the su(3) limit of IBM 
is a formidable problem, since the su$_q$(3) $\supset$ so$_q$(3) 
decomposition has for the moment been achieved only for completely symmetric 
su$_q$(3) irreducible representations.

6) A $q$-deformed version of the Moszkowski model 
as well as a $q$-deformed Moszkowski model 
with cranking have been developed. The 
possibility of using $q$-deformation in assimilating temperature effects is 
receiving attention, since it has also been found 
that this approach
can be used in describing thermal effects in the framework of a $q$-deformed 
Thouless model for supercoductivity. 
In addition, $q$-deformed versions of the Lipkin-Meshkov-Glick (LMG) 
model have been developed, both for the 2-level version of the model
in terms of an su$_q$(2) algebra, and for the 3-level version 
of the model in terms of an su$_q$(3) algebra. 

7) It has been proved that a generalized deformed u(2)
algebra is the symmetry algebra of the two-dimensional anisotropic quantum
harmonic oscillator with rational ratios of frequencies (RHO), 
which is the oscillator describing the 
single-particle
level spectrum of ``pancake'' nuclei, i.e. of very oblate 
triaxially deformed nuclei
with $\omega_x >> \omega_y$, $\omega_z$. Furthermore, a generalized 
deformed u(3) algebra turns out to be the symmetry algebra of the 
three-dimensional RHO, which is related to the symmetry underlying 
the structure of superdeformed and hyperdeformed nuclei.  

8) Recently the 3-dimensional $q$-deformed (isotropic) harmonic oscillator
has been studied in detail [4]. 
It turns out that in this framework, one can reproduce 
level schemes similar to the ones occuring in the modified harmonic 
oscillator model, first suggested by Nilsson. 
An appropriate $q$-deformed spin--orbit interaction term has also been
developed [4]. Including this term in the 3-dimensional 
$q$-deformed (isotropic) harmonic oscillator scheme one can reproduce 
level schemes similar to these provided by the modified harmonic 
oscillator with spin--orbit interaction. 

\vskip 28 truept
\centerline{\bf 3. QUANTUM ALGEBRAS IN MOLECULAR STRUCTURE}
\vskip 12 truept 

Similar techniques can be applied
in describing properties of diatomic and polyatomic molecules. A brief
list will be given here. More details and relevant references can be found 
in the review articles [1--3]. 

1) Rotational spectra of diatomic molecules have been described in terms of 
the su$_q$(2) model.  As in the case of nuclei, $q$ is a phase factor 
($q=e^{i\tau}$). In molecules $\tau$ is of the order of 0.01. 
The use of the su$_q$(2) symmetry leads to a partial summation of the Dunham
expansion describing the rotational--vibrational spectra of diatomic 
molecules. Molecular backbending (bandcrossing) has also been
described in this framework. Rotational spectra of symmetric top molecules 
have also been considered in the framework of the su$_q$(2) symmetry. 
Furthermore, two $q$-deformed rotators with slightly different parameter 
values have been used  for the description of $\Delta I=1$ staggering
effects in rotational bands of diatomic molecules. 

2) Vibrational spectra of diatomic molecules have been described in terms of 
$q$-deformed anharmonic oscillators having the su$_q$(1,1)
or the u$_q$(2) $\supset$ o$_q$(2)
symmetry, as well as in terms of generalized deformed oscillators.
These results, combined with 1), lead 
to the full summation of the Dunham expansion. 
A two-parameter deformed anharmonic oscillator with u$_{qp}$(2) $\supset$ 
o$_{qp}$(2) symmetry has also been considered. 

3) The physical content of the anharmonic oscillators mentioned in 2) 
has been clarified by constructing WKB equivalent potentials (WKB-EPs) 
and classical equivalent potentials providing approximately the same
spectrum. 
The results have been corroborated by the study of the 
relation between su$_q$(1,1) and the anharmonic oscillator with  $x^4$ 
anharmonicities. Furthermore 
the WKB-EP corresponding to the su$_q$(1,1) anharmonic 
oscillator has been connected to a class of Quasi-Exactly Soluble Potentials 
(QESPs).                                                         

4) Generalized deformed oscillators giving the same spectrum as the Morse 
potential and the modified P\"oschl--Teller potential,  
as well as a deformed oscillator containing them as special cases  
have also been constructed. 
In addition,  $q$-deformed versions of the Morse potential have been given, 
either by using the so$_q$(2,1) symmetry or by solving a 
$q$-deformed Schr\"odinger equation for the usual Morse potential. 
For the sake of completeness it should be mentioned that 
a deformed oscillator giving the same spectrum as the Coulomb 
potential has also been constructed.  

5) A $q$-deformed version of the vibron model for diatomic molecules has been 
constructed. 

6) For vibrational spectra of polyatomic molecules a model of $n$ coupled 
generalized deformed oscillators has been built, containing the 
approach of Iachello and Oss as a special case. In addition
a model of two $Q$-deformed oscillators coupled so that the total 
Hamiltonian has the su$_Q$(2) symmetry has been proved to be 
equivalent, to lowest order approximation, to a system of two identical Morse
oscillators coupled by the cross-anharmonicity usually used 
empirically in describing vibrational spectra of diatomic molecules.  

7) Quasi-molecular resonances in the systems $^{12}$C+$^{12}$C and 
$^{12}$C+$^{16}$O have been described in terms of a $q$-deformed oscillator
plus a rigid rotator.

\vskip 28 truept
\centerline{\bf 4. QUANTUM ALGEBRAIC SYMMETRIES IN ATOMIC CLUSTERS} 
\vskip 12 truept 
\noindent
{\it 4.1 Introduction}
\vskip  12 truept

In this Section we will discuss in some detail an application of quantum 
algebraic techniques to atomic clusters. Because of lack of space, no detailed
list of references will be given. For the relevant references the reader 
is referred to Refs [4--6]. 

Metal clusters have been recently the subject of many investigations.
One of the first 
fascinating findings in their study was the appearance of magic numbers, 
analogous to 
but different from the magic numbers appearing in the shell structure of 
atomic nuclei. Different kinds of metallic clusters [alkali
metals (Na, Li, K, Rb, Cs),  
noble metals (Cu, Ag, Au),  
divalent metals of the IIB group (Zn, Cd), 
trivalent metals of the III group (Al, In)] 
exhibit different sets of magic numbers. 
The analogy between the magic numbers observed in metal clusters and the 
magic numbers observed in atomic nuclei 
led to the early description of metal 
clusters in terms of the Nilsson--Clemenger model,
which is a simplified version of the Nilsson model of atomic 
nuclei, in which no spin-orbit interaction is included. Further theoretical
investigations in terms of the jellium model  
demonstrated that the mean field potential in the case of simple metal 
clusters bears great similarities to the Woods--Saxon potential 
of atomic nuclei, with a slight modification of the ``wine bottle''
type. The Woods--Saxon potential itself looks like a harmonic 
oscillator truncated at a certain energy value and flattened at the bottom. 
It should also be recalled that an early schematic explanation of the 
magic numbers of metallic clusters has been given in terms of a scheme 
intermediate between the level scheme of the 3-dimensional harmonic 
oscillator and the square well. Again in this case the 
intermediate 
potential resembles a harmonic oscillator flattened at the bottom.  

On the other hand, modified versions of harmonic oscillators have been 
recently investigated in the novel mathematical framework of quantum
algebras, 
which are nonlinear generalizations of the usual Lie
algebras. The spectra of $q$-deformed oscillators increase either 
less rapidly (for $q$ being a phase factor, i.e. $q=e^{i\tau}$ with 
$\tau$ being real) or more rapidly (for $q$ being real, i.e. $q=e^{\tau}$ 
with $\tau$ being real)
in comparison to the equidistant spectrum 
of the usual harmonic oscillator, while the corresponding 
(WKB-equivalent) potentials
resemble the harmonic oscillator potential,
truncated at a certain energy (for $q$ being a phase factor) 
or not (for $q$ being real), 
the deformation inflicting an overall
widening or narrowing of the potential, depending on the value of the 
deformation parameter $q$.   

Very recently, a $q$-deformed version of the 3-dimensional harmonic 
oscillator has been constructed [4], taking advantage of the 
u$_q$(3) $\supset$ so$_q$(3) symmetry. 
The spectrum of this 3-dimensional $q$-deformed harmonic oscillator 
has been found [4] to reproduce very well the spectrum of the 
modified harmonic oscillator introduced by Nilsson, without the 
spin-orbit interaction term. Since the Nilsson model without the 
spin orbit term is essentially the Nilsson--Clemenger model used 
for the description of metallic clusters, it is worth examining 
if the 3-dimensional $q$-deformed harmonic oscillator can reproduce 
the magic numbers of simple metallic clusters. This is the subject 
of the present Section. 

%In Subsection 4.2 the 3-dimensional $q$-deformed harmonic oscillator will
%be briefly described, while in Subsection 4.3 the magic numbers provided 
%by this oscillator will be compared with the experimental data for Na and Li
%clusters, as well as with the predictions of other theories (jellium model, 
%Woods--Saxon and wine bottle potentials, classification scheme using the 
%$3n+l$ pseudo quantum number). Additional comparisons of magic numbers 
%predicted by the 3-dimensional $q$-deformed harmonic oscillator to 
%experimental data and to the results of other theoretical approaches 
%will be made in Subsection 4.4 (for other alkali metal clusters 
% and noble metal clusters), Subsection 4.5 (for divalent group IIB metal 
%clusters), and Subsection 4.6 (for 
%trivalent group III metal clusters), while Subsection 4.7 will contain
%discussion of the present results and plans for further work. 

\vskip 28 truept
\noindent
{\it 4.2 The 3-dimensional $q$-deformed harmonic oscillator} 
\vskip 12 truept

The space of the 3-dimensional $q$-deformed harmonic oscillator consists of 
the completely symmetric irreducible representations of the quantum algebra
u$_q$(3). In this space a deformed angular momentum algebra, so$_q$(3), 
can be defined [4]. The Hamiltonian of the 3-dimensional $q$-deformed 
harmonic oscillator is defined so that it satisfies the following 
requirements:

a) It is an so$_q$(3) scalar, i.e. the energy is simultaneously measurable
with the $q$-deformed  angular momentum related to the algebra so$_q$(3) 
and its $z$-projection.   

b) It conserves the number of bosons, in terms of which the quantum 
algebras u$_q$(3) and so$_q$(3) are realized. 

c) In the limit $q\to 1$ it is in agreement with the Hamiltonian of the usual 
3-dimensional harmonic oscillator. 
 
It has been proved [4] that the Hamiltonian of the 3-dimensional 
$q$-deformed harmonic oscillator satisfying the above requirements 
takes the form
$$ H_q = \hbar \omega_0 \left\{ [N] q^{N+1} - {q(q-q^{-1})\over [2] }
C_q^{(2)}
\right\}, \eqno(1)$$
where $N$ is the number operator and $C_q^{(2)}$ is the second order 
Casimir operator of the algebra so$_q$(3), while 
$$ [x]= {q^x-q^{-x} \over q-q^{-1}} \eqno(2) $$
is the definition of $q$-numbers and $q$-operators. 

The energy eigenvalues of the 3-dimensional $q$-deformed harmonic oscillator 
are then [4]
$$ E_q(n,l)= \hbar \omega_0 \left\{ [n] q^{n+1} - {q(q-q^{-1}) \over [2]}
[l] [l+1] \right\}, \eqno(3)$$
where $n$ is the number of vibrational quanta and $l$ is the eigenvalue of
the 
angular momentum, obtaining the values
$l=n, n-2, \ldots, 0$ or 1.  

In the limit of $q\to 1$ one obtains ${\rm lim}_{q\to 1} E_q(n,l)=
\hbar \omega_0 n$, which coincides with the classical result. 

For small values of the deformation parameter $\tau$ (where $q=e^{\tau}$)
one can expand Eq. (3) in powers of $\tau$  obtaining [4]
$$E_q(n,l)= \hbar \omega_0 n -\hbar \omega_0 \tau \left(l(l+1)-n(n+1)\right)$$
$$ -\hbar \omega_0 \tau^2 \left( l(l+1)-{1\over 3} n(n+1)(2n+1) \right)
+ {\cal O} (\tau^3).\eqno(4)$$

The last expression to leading order bears great similarity to the modified 
harmonic oscillator suggested by Nilsson 
(with the spin-orbit term omitted)
$$ V= {1 \over 2} \hbar \omega \rho^2 -\hbar \omega \kappa' 
({\bf L}^2 - <{\bf L}^2>_N ), \qquad \rho=r \sqrt {M\omega \over \hbar}
,\qquad
<{\bf L}^2>_N = {N(N+3)\over 2}.\eqno(5)$$ 
The energy eigenvalues of Nilsson's  modified harmonic oscillator are 
$$ E_{nl}= \hbar \omega n -\hbar \omega \mu' \left( l(l+1)-{1\over 2}
n(n+3)\right). \eqno(6) $$
It has been proved [4]  that the spectrum of the 3-dimensional 
$q$-deformed harmonic oscillator closely reproduces the spectrum of 
the modified harmonic oscillator of Nilsson. In both cases the effect 
of the $l(l+1)$ term is to flatten the bottom of the harmonic 
oscillator potential, thus making it to resemble the Woods--Saxon 
potential. 

In the spectrum of the 3-dimensional $q$-deformed harmonic oscillator 
each level is characterized by the quantum numbers 
$n$ (number of vibrational quanta) and $l$ (angular momentum). 
The number of particles which can be accommodated by a level with angular 
momentum $l$ is equal to $2(2l+1)$). The total number of particles up to 
and including this level is given by the sum of the quantity $2(2l+1)$ 
for all levels up to and including the level in question. If the energy 
difference between two successive levels is larger than 
a given number, which for future reference we call $\delta$, 
 it is  considered as a gap separating two successive shells, so that 
the number of particles which can be accommodated up to the gap is a 
magic number. If the energy separation is smaller than $\delta$, the
successive levels are considered as belonging to the same shell and no 
magic number occurs at this point. 

In Table 1 of Ref. [5] and in Tables 1, 2, 3 of Ref. [6] one can find various 
level schemes of the 
3-dimensional $q$-deformed harmonic oscillator for various values 
of the deformation parameter ($\tau = 0.020$, 0.038, 0.050) and the 
energy gap ($\delta = 0.20$, 0.26, 0.38, 0.39).
We remark that the small magic numbers 
do not change much as the parameter $\tau$ is varied, while large magic 
numbers get more influenced by the parameter modification. 

\vskip 28 truept

\noindent{\it 4.3 Sodium and lithium clusters}
\vskip 12 truept

The magic numbers provided by the 3-dimensional $q$-deformed harmonic 
oscillator with $\tau=0.038$ and $\delta = 0.39$ 
have been compared to available experimental data for 
Na clusters and Li clusters 
in Table 4 of Ref. [6]. The following comments apply:

i) Only magic numbers up to 1500
are reported, since it is known that filling of electronic shells 
is expected to occur only up to this limit. For large 
clusters beyond this point it is known that magic numbers can be explained by
the completion of icosahedral or cuboctahedral shells of atoms. 

ii) Up to 600 particles there is consistency among the various experiments 
and between the experimental results in one hand and our findings in the 
other. 

iii) Beyond 600 particles the results of the four  experiments,
which report magic numbers in this region, are 
quite different. However, the results of all four  experiments are 
well accommodated by the present model. In addition, each magic number 
predicted by the model is supported by at least one experiment. 

In Table 4 of Ref. [6] 
the predictions of three simple theoretical models 
(non-deformed 3-dimensional harmonic oscillator, square well potential, 
rounded square well potential (intermediate between the 
previous two)) are also reported for comparison. It is clear 
that the predictions of the non-deformed 3-dimensional harmonic oscillator
are 
in agreement with the experimental data only up to magic number 40, 
while the other two models give correctly a few more magic numbers (58, 
92, 138), although they already fail by predicting magic numbers at 68, 70, 
106, 112, 156, which are not observed.  

It should be noticed at this point that the first few magic numbers of 
alkali clusters (up to 92) can be correctly reproduced by the assumption 
of the formation of shells of atoms instead of shells of delocalized 
electrons, this assumption being applicable  under conditions 
not favoring delocalization of the valence electrons of alkali atoms. 

Comparisons among the present results, experimental data for Na and Li 
clusters, and
theoretical predictions more sophisticated than these reported in Table 4
of Ref. [6], have been made in Table 5 of Ref. [6], where magic numbers 
predicted by various jellium model calculations, Woods--Saxon and wine bottle 
potentials, as well as by a classification scheme using the $3n+l$ pseudo 
quantum number are reported. The following observations can be made:

i) All magic numbers predicted by the 3-dimensional $q$-deformed harmonic 
oscillator are supported by at least one experiment, with no exception.

ii) Some of the jellium models, as well as the $3n+l$ classification scheme, 
predict magic numbers at 186, 540/542, which are not supported by 
experiment. Some jellium models also predict a magic number at 
748 or 758, again without support from experiment. Woods--Saxon 
and wine bottle potentials predict a magic number at 
68, for which no experimental support exists. The present scheme 
avoids problems at these numbers. It should be noticed, however, 
that in the cases of 186 and 542 the energy gap following them 
in the present scheme is 
0.329 and 0.325 respectively (see Table 1 of Ref. [6]), i.e. quite close to 
the threshold of 0.39 which we have considered as the minimum energy 
gap separating different shells. One could therefore qualitatively 
remark that 186 and 542 are ``built in'' the present scheme as
``secondary'' (not very pronounced) magic numbers.  

\vfill\eject
\vskip 28 truept
\noindent
{\it 4.4 Other alkali metals and noble metals} 
\vskip 12 truept

Experimental data for various alkali metal clusters (Li, Na, K, Rb, Cs) and 
noble metal clusters (Cu, Ag, Au) 
have been reported in Table 6 of Ref. [6], along with the 
theoretical predictions of the 3-dimensional $q$-deformed harmonic oscillator 
with $\tau=0.038$ and $\delta =0.39$. The following comments apply:

i) In the cases of Rb, Cu, Ag, and 
Au, what is seen experimentally is cations 
of the type Rb$^+_N$, Cu$^+_N$, Ag$^+_N$, Au$^+_N$, which contain $N$ atoms 
each, but $N-1$ electrons. The magic numbers reported in  Table 6 of Ref. [6] 
are electron magic numbers in all cases. 

ii) All alkali metals and noble metals give the same magic numbers, at least 
within the ranges reported in the table. For most of these metals the 
range of experimentally determined magic numbers is rather limited, 
with Na, Cs, Li, and Ag being notable exceptions. 

iii) The magic numbers occuring in Na, Cs, Li, and Ag 
are almost identical, and are 
described very well by the 3-dimensional $q$-deformed harmonic oscillator 
with $\tau = 0.038$ and $\delta =0.39$. 
The limited data on K, Rb, Cu, Au, also agree with the magic 
numbers of the same oscillator.  

\vskip 28 truept
\noindent
{\it 4.5 Divalent metals of the IIB group}
\vskip 12 truept

For these metals the quantities determined experimentally are 
numbers of atoms exhibiting ``magic''
behaviour. Each atom has two valence electrons, therefore the magic numbers 
of electrons are twice the magic numbers of atoms. The magic numbers of 
electrons for Zn and Cd clusters have been reported in Table 7
of Ref. [6], along 
with the magic numbers predicted by the 3-dimensional $q$-deformed harmonic 
oscillator for two different parameter values 
($\tau=0.038$, $\delta =0.26$ and $\tau=0.020$, $\delta =0.20$), 
and the magic numbers given by a potential intermediate between the simple
harmonic oscillator and the square well potential. 
The following comments can be made:

i) The experimental magic numbers for Zn and Cd are almost 
identical. Magic numbers reported in parentheses are ``secondary'' magic 
numbers, while the magic numbers without parentheses are the ``main'' ones, 
as indicated in the experimental papers. 

ii) In Table 7 o Ref. [6] magic numbers of the 3-dimensional 
$q$-deformed 
harmonic oscillator with $\tau=0.038$ and energy gaps larger than 0.26 
are reported. Decreasing the energy gap considered as separating different 
shells from 0.39 (used in Table 1 of Ref. [6]) to 0.26 (used in Table 7
of Ref. [6]) has as a 
result that the numbers 70 and 106 become magic, in close agreement with 
the experimental data. Similar but even better results are gotten from 
the 3-dimensional $q$-deformed harmonic oscillator 
 characterized by $\tau =0.020$, 
with the energy gap between different shells being set equal to 0.20~. 
We observe that the second oscillator predicts an additional magic number 
at 112, in agreement with experiment, 
but otherwise gives the same results as the first one. 
We remark therefore that the general agreement between the results given 
by the 3-dimensional $q$-deformed harmonic oscillator and the experimental 
data is not sensitively dependent on the parameter value, but, in contrast, 
quite different parameter values ($\tau=0.038$, $\tau=0.020$) provide 
quite similar sets of magic numbers (at least in the region of relatively 
small magic numbers). 

iii) Both oscillators reproduce all the ``main'' magic numbers of Zn and Cd,
while the intermediate potential between the simple harmonic oscillator and 
the square well potential, reported in the same table,  reproduces all the
 ``main'' magic numbers except 108. 

\vskip 28 truept
\noindent
{\it 4.6 Trivalent metals of the III group} 
\vskip 12 truept

Magic numbers of electrons for the trivalent metals Al and In 
are reported in Table 7 of Ref. [6], along with the 
predictions of the 3-dimensional $q$-deformed harmonic oscillator 
with $\tau=0.050$ and $\delta = 0.38$. The following comments can be made: 

i) It is known  that small magic numbers in 
clusters of Al and In cannot be explained by models based on the filling 
of electronic shells, because of symmetry breaking caused by the ionic 
lattice, while for large magic numbers this problem  does not exist. 

ii) The 3-dimensional $q$-deformed harmonic oscillator 
with $\tau=0.050$ and $\delta =0.38$ provides magic numbers 
which agree quite well with the experimental findings, with an 
exception in the region of small magic numbers, where the model
fails to reproduce 
the magic numbers 164 and 198, predicting only a magic number at 186.
In addition the oscillator predicts magic numbers at 398, 890, 1074, 
which are not seen in the experiment.

\vskip 28 truept
\noindent
{\it 4.7 Discussion} 
\vskip 12 truept

The following general remarks can now be made:

i) From the results reported above it is quite clear that the 
3-dimensional $q$-deformed harmonic oscillator describes very well 
the magic numbers of alkali metal clusters and noble metal clusters 
in all regions, using only one free parameter ($q=e^{\tau}$ with 
$\tau=0.038$). It also provides an accurate description of the 
``main'' magic numbers of clusters of divalent group IIB metals,
either with the same parameter value ($\tau=0.038$) or with 
a different one ($\tau=0.020$). In addition it gives a satisfactory 
description of the magic numbers of clusters of trivalent group III metals
with a different parameter value ($\tau=0.050$). 

ii) It is quite remarkable that the 3-dimensional $q$-deformed harmonic 
oscillator reproduces long sequences of
magic numbers (Na, Cs, Li, Ag) at least as accurately as other,
more sophisticated, models by using only one free parameter ($q=e^{\tau}$). 
Once the parameter is fixed, the whole spectrum is fixed and no further 
manipulations can be made.
This can be considered as evidence that the 3-dimensional $q$-deformed 
harmonic oscillator owns a symmetry (the u$_q$(3) $\supset$ so$_q$(3)
symmetry) appropriate for the description of the physical systems under 
study. 

iii) It has been remarked that if $n$ is the number of nodes 
in the solution of the radial Schr\"odinger equation and $l$ is the 
angular momentum quantum number, then the degeneracy of energy levels of 
the hydrogen atom characterized by the same $n+l$ is due to the so(4) 
symmetry of this system, while the degeneracy of energy levels of the 
spherical harmonic oscillator (i.e. of the 3-dimensional isotropic 
harmonic oscillator) characterized by the same $2n+l$ 
is due to the su(3) symmetry of this system. $3n+l$ has been used 
to approximate the magic numbers of alkali metal clusters
with some success, but no relevant Lie symmetry could be determined. In view
of the present findings the lack of Lie symmetry related to $3n+l$ is quite 
clear: the symmetry of the system appears to be a quantum algebraic 
symmetry (u$_q$(3)), which is a nonlinear extension of the Lie 
symmetry u(3). 

iv) An interesting problem is to determine a WKB-equivalent potential 
giving (within this approximation) the same spectrum as the 
3-dimensional $q$-deformed harmonic oscillator, using methods similar 
to these  of Refs [1,2]. The similarity
between the results of the present model and these provided by the 
Woods--Saxon potential (Table 5 of Ref. [6])  suggests that the answer 
should be a harmonic oscillator potential flattened at the bottom, 
similar to the Woods--Saxon potential. If such a WKB-equivalent 
potential will show any similarity to a wine bottle shape,
as several potentials used for the description of metal clusters do, 
remains to be seen. 

In summary, we have shown that the 3-dimensional $q$-deformed harmonic 
oscillator with u$_q$(3) $\supset$ so$_q$(3) symmetry correctly 
predicts all experimentally observed magic numbers of alkali metal clusters
and of noble metal clusters   
up to 1500, which is the expected limit of validity for theories based on 
the filling of electronic shells. In addition it gives a good description 
of the ``main'' magic numbers of group IIB (divalent) metal clusters, 
as well as a satisfactory description of group III (trivalent) metal
clusters. 
This indicates that u$_q$(3), which 
is a nonlinear deformation of the u(3) symmetry of the spherical
(3-dimensional isotropic) harmonic oscillator, is a good candidate for 
being the symmetry of systems of several  metal clusters.  

\vskip 28 truept 
\centerline{\bf REFERENCES} 
\vskip 12 truept 

\item{[1]}
D. Bonatsos and C. Daskaloyannis, {\it Progr. Part. Nucl. Phys.} 
{\bf 43} (1999) in press. nucl-th/9909003. 

\item{[2]} 
D. Bonatsos, C. Daskaloyannis, P. Kolokotronis and D. Lenis, 
{\it Rom. J. Phys.} {\bf 41}, 109 (1996). nucl-th/9512017. 

\item{[3]} 
P. P. Raychev, {\it Adv. Quant. Chem.} {\bf 26}, 239 (1995). 

\item{[4]} 
P. P. Raychev, R. P. Roussev, N. Lo Iudice and P. A. Terziev, 
{\it J. Phys. G} {\bf 24}, 1931 (1998). 

\item{[5]}
D. Bonatsos, N. Karoussos, P. P. Raychev, R. P. Roussev and P. A. Terziev, 
{\it Chem. Phys. Lett.} {\bf 302}, 392 (1999). quant-ph/9909002. 

\item{[6]}
D. Bonatsos, N. Karoussos, P. P. Raychev, R. P. Roussev and P. A. Terziev, 
NCSR Demokritos preprint DEM-NT-99-07 (1999). 

\vfill\eject\bye